\documentclass{article}

\usepackage{arxiv}

\usepackage[utf8]{inputenc} 
\usepackage[T1]{fontenc}    
\usepackage{hyperref}       
\usepackage{url}            
\usepackage{booktabs}       
\usepackage{amsmath,amssymb,amsfonts}       
\usepackage{nicefrac}       
\usepackage{microtype}      
\usepackage{lipsum}		    
\usepackage{graphicx}
\usepackage{natbib}
\usepackage{doi}
\usepackage{multirow}%
\usepackage{amsthm}%
\usepackage{mathrsfs}%
\usepackage[title]{appendix}%
\usepackage{xcolor}%
\usepackage{textcomp}%
\usepackage{manyfoot}%
\usepackage{algorithm}%
\usepackage{algorithmicx}%
\usepackage{algpseudocode}%
\usepackage{listings}%

\title{Accounting for Preferential Sampling Using a Constructed Covariate}

\author{
	\href{https://orcid.org/0000-0001-9025-2962}{Andreia~Monteiro}\thanks{Corresponding author.}\\
	Centre of Mathematics of the University of Minho (CMAT), Braga, Portugal\\
	\texttt{andreiaforte50@gmail.com}
	\And
	\href{https://orcid.org/0000-0001-6020-9373}{Isabel~Nat\'{a}rio}\\
	Department of Mathematics of the Nova School of Science and Technology\\
	Center for Mathematics and Applications (NOVA Math)\\
	NOVA University of Lisbon, Caparica, Portugal\\
	\texttt{icn@fct.unl.pt}
	\And
	\href{https://orcid.org/0000-0003-4905-8134}{Ivone~ Figueiredo}\\
	Portuguese Institute for Sea and Atmosphere (IPMA)\\
	Center of Statistics and its Applications (CEAUL), Lisbon, Portugal\\
	\texttt{ifigueiredo@ipma.pt}
		\And
	\href{https://orcid.org/0000-0002-1074-1297}{Paula~ Simões}\\
Military Academy Research Center - Military University Institute (CINAMIL)\\
Engineering Superior Institute of Lisbon (ISEL), Polytechnic Institute of Lisbon\\
Centre for Mathematics and Applications (NOVA Math)\\
 NOVA University of Lisbon, Portugal\\
	\texttt{pc.simoes@fct.unl.pt}
}

\renewcommand{\shorttitle}{Accounting PS using a constructed covariate}

\hypersetup{
pdftitle={A template for the arxiv style},
pdfsubject={q-bio.NC, q-bio.QM},
pdfauthor={David S.~Hippocampus, Elias D.~Striatum},
pdfkeywords={First keyword, Second keyword, More},
}

\begin{document}
\maketitle

\begin{abstract}
{In geostatistics, it is commonly assumed that sampling locations are selected independently of the underlying spatial process. In practice, however, this assumption is frequently violated. In fisheries, for example, sampling sites are often chosen to maximize expected catches, creating a stochastic dependence between the abundance process and the sampling design. Such preferential sampling can introduce substantial bias and compromise statistical inference. This study investigates the use of constructed covariates, based on average distances from nearest neighbours observations, that are able to mitigate preferential sampling. The inclusion of such covariate in the geostatistical model might be able to account for the stochastic dependence of sampling locations on the spatial variable. If this inclusion sufficiently captures the dependence, conventional methods of inference may be applied without resorting to more complex models. The proposed methodology is evaluated through an extensive simulation study that explores a variety of sampling scenarios and spatial configurations. Additionally, we demonstrate the practical utility of the approach using two real-world datasets: one on fishery landings provided by the Instituto Português do Mar e da Atmosfera, and another concerning lead pollution biomonitoring in Galicia.  Results show that incorporating the constructed covariate can substantially reduce the impact of preferential sampling, enabling reliable inference with standard geostatistical tools. We also discuss practical challenges, limitations, and paths for future methodological development.}

\end{abstract}

\keywords {Preferential sampling  \and Constructed covariates  \and Nearest neighbour distances  \and Geostatistics}

\section{Introduction}
\label{intro}

Geostatistical methods are used to model continuous spatial phenomena of interest, described by a spatially continuous stochastic process, observed with error in sampling locations that are either considered as fixed or stochastically independent of the underlying process, \cite{diggle2007}.

However, in practice, the choice of sampling locations in a spatial network is often guided by budget and other practical requirements. For example, it is well-known that in air pollution studies the monitors are typically placed near the most likely pollution sources and in areas of higher population density, \cite{zidek2014}. In marine based species distribution models, \cite{robinsonetal2017}, or in fish abundance studies, \cite{frenchetal2021}, frequently relying on fishery data that can only be observed when and where the resource is available, sampling locations are deliberately chosen guided by a belief regarding the abundance of the species of interest, \cite{pennino2018}. In these cases, the aforementioned assumption of independence fails since the process under study determines data-locations.

This problem, coined preferential sampling in the context of spatial statistics, occurs, if a stochastic dependence exists between the choice of sampling locations and the underlying process being measured. Interest in preferential sampling was sparked in 2010 by the landmark paper by \cite{diggle2010}. The authors demonstrated that the consequences of preferential sampling on spatial inference can be severe, with both spatial prediction and parameter
estimation affected. In the same way \cite{pennino2018} showed that ignoring the preferential nature of the sampling process can lead to incorrect estimates and large biases in the prediction of the underlying spatial process. \cite{watson2020fast} claims that predictions of the process can be severely biased when standard statistical methodologies are applied to preferentially sampled data without adjustment.

Since that landmark paper, preferential sampling has been identified as a major concern
across multiple fields including environmental statistics, ecology and econometrics. In species richness studies, preferential sampling may occur due to data being comprised of opportunistic sightings. Observers frequently focus their efforts in areas where they expect to find the species, \cite{watson2020fast} . \cite{conn2017} use geostatistical methods to model ecological data obtained by preferential sampling, referring to a special case of opportunistic sampling in which there is stochastic dependence between the sampling design and the reported species counts. \cite{pennino2018} present an approach for modeling the distribution of species using opportunistic data and show that predictive maps significantly improve the prediction of the target species when the model accounts for preferential sampling.

The mitigation of the preferential sampling effects has been accomplished through its direct modelling within the Geostatistical model based approach of \cite{diggle1998}. The idea, also identified as a marked point process for the locations, \cite{stoyan2008}, is to joint model the observation process (marks) and the sampling process (points), \cite{diggle2010}, an exercise that might be computationally challenging.

Even though some of the computational difficulties of fitting joint models have been smoothed by framing these problems within Bayesian inference, \cite{pati2011}, detecting preferential sampling or dependence between marks and points in a point process approach is therefore an important issue, \cite{gelfand2012}, specially if something can be done to avoid the joint modelling process. This is sometimes the case when there are covariates available that, when included in the model, seem to be sufficient to account for this relationship between points and marks. The discovery and inclusion of these covariates in the model may just be enough to justify the use of standard methodologies for doing the modelling, \cite{watson2021}.

Based on this thought, the motivation of this work is to develop and include a constructed covariate in the observation process model that is able to account for the local structure of the point pattern of the sampling process. This is in line, but is a different approach,  with the work of \cite{illian2012toolbox}, \cite{illian2013} and \cite{raeisi2020}, that also consider a similar kind of constructed covariate, including it in the intensity of a Cox process model for the points process rather than in the observation process model, measuring the local structure of the point pattern associated with an additional spatial effect at medium-long range. They considered constructed covariates based on first order summary characteristics defined for any location in the observation window, unlike our approach that only defines the covariate for the observations locations.

We investigate the use of constructed covariates that are able to explain preferential sampling. For that, we consider a number of different proposals, based on average distances between each observation point and the other observations, chosen according to different criteria: the $k$ nearest points to all other observation points and those points that are less than a fixed distance $d$ apart. The inclusion of this covariate is intended to account for the dependence between the marks and the point locations by capturing the small-scale spatial variation of the points, thereby eliminating the need to specify a more complex marked point process and enabling inference based on a single model for the marks. If, after including this covariate, no residual evidence of dependence between the marks and the point locations is detected, inference can be based on standard statistical techniques, considerably simplifying the modelling problem.

The remainder of the paper is structured as follows. Section 2 introduces the standard geostatistical framework for modelling preferential sampling and outlines its key assumptions and limitations. In Section 3, we present several formulations of the constructed covariates, based on average distances to nearest neighbours sampling locations, designed to capture the spatial dependence between the sampling process and the underlying variable of interest.
Section 4 contains an extensive simulation study, carefully designed to assess the performance and robustness of the proposed methodology under a variety of spatial configurations and sampling intensities. In Section 5, we illustrate the practical application of our approach using two real-world datasets: the first concerns black scabbardfish catch records provided by the Instituto Português do Mar e da Atmosfera (IPMA), covering the South fishing grounds of Portugal from 2009 to 2013; the second relates to lead pollution in Galicia, northern Spain, with concentrations measured through moss biomonitoring, as originally reported in \cite{diggle2017}.
Finally, Section 6 offers concluding remarks and outlines potential avenues for future research.

\section{Geostatistical model for Preferential Sampling }

The term geostatistics identifies the part of spatial statistics which is concerned with
data obtained by spatially discrete sampling of a spatially continuous process $\left\lbrace S(x) : x \in \mathbb{R}^2\right\rbrace $. In Geostatistics, we can have regular or irregular sampling but, the usual assumption is that the selection of the sampling locations does not depend on the values of the spatial variable. In fact, most methods are based on the assumption, possibly tacit, that sampling locations are uniformly distributed over the observed region. However, there are situations in which the process under study determines the data-locations and the above mentioned assumption is violated, being in presence of preferential sampling: the sampling process and the observed process are dependent and there is an underlying stochastic relationship between data and locations. \cite{diggle2010}
defines preferential sampling succinctly to any situation in which $[S,X]\neq [S][X]$, here $S$ denotes the spatial field and $X$ the locations, $[\centerdot]$ means the ``distribution of''.

\cite{diggle2010} developed a model for geostatistical data collected in a preferential way, where sampling locations and observations are jointly modelled depending on a common unobserved spatial random field. According to authors, the model for the observations takes the form
\begin{equation} \label{eq_geo}
	Y(x_i)=S(x_i)+ W_i,
\end{equation}
$Y(x_i)$ denotes the measured value at location $x_i$, $S$ is a stationary Gaussian Process with mean $\mu_s$, marginal variance $\sigma_s^2$ and Matérn correlation function with shape parameter $\kappa$ and scale (range) parameter $\phi$. $W_i$ are Gaussian random errors with mean 0, variance $\tau^2$, $i=1 \cdots n$, with $n$ the number of observations.

The model for the locations of the observation is a log Gaussian Cox process with intensity
\begin{equation}\label{eq_lambda}
	\Lambda(x_i)=\exp\left\lbrace \alpha + \beta S(x_i) \right\rbrace,
\end{equation}
where parameter $\beta$ controls the degree of preferentiality, for example, when $\beta >0$ the sample points are concentrated predominantly near higher values of the spatial random field and
when $\beta=0$ that corresponds to the situation of a non-preferential sampling, modelled by an homogeneous Poisson process with intensity $\exp(\alpha)$.\\

The modelling approach suggested by \cite{diggle2010}, accounts for preferential sampling using likelihood-based inference with Monte Carlo methods, however the convergence of this algorithm is very slow and the running time becomes burdensome for larger datasets and a large number of Monte Carlo samples. Bayesian inference based on a SPDE-INLA approach has more recently been used, \cite{dinsdale2019}, with considerable advantages on the computational burden. The aforementioned joint geostatistical model can be regarded as a marked point process, the marks modelling the observed quantities and the points the sampling locations. This relationship allows the use of methodologies developed in the context of point processes for the analysis of preferential sampling, \cite{monteiro2022}.

It is worth noting that, despite recent methodological advances, joint models for preferential sampling remain challenging in practice. They are not yet widely available in mainstream statistical software, often require substantial computational resources, and can be complex to specify and interpret. As a result, these models may remain inaccessible to many researchers, particularly those without specialized expertise in spatial statistics or advanced computational tools.

\section{Constructed Covariates}

Given the substantial impact of preferential sampling on statistical inference and the practical challenges associated with fitting joint models, both in terms of computational burden and implementation complexity, detecting the presence of preferential sampling becomes a critical step in spatial data analysis. In particular, identifying dependence between the sampling locations (points) and the observed values (marks), as framed in the point process perspective, is essential to ensure valid inference.

When preferential sampling is detected, the availability and appropriate inclusion of covariates in the geostatistical model (\ref{eq_geo}) can play a pivotal role. If such covariates sufficiently capture the underlying dependence between the sampling design and the spatial process, they may effectively mitigate the bias introduced by preferential sampling. In such cases, standard modelling approaches may remain appropriate, offering a simpler and more accessible alternative to complex joint models. Thus, careful exploration, selection, and incorporation of relevant covariates is not only desirable but may be crucial for reliable and interpretable spatial analysis \cite{watson2021}.

In this paper we propose an approach to address the preferential sampling issue including in the observation model a constructed covariate that is able to mitigate this effect, that is able to account for the local structure of the point pattern of the sampling process, explaining preferential sampling. Constructed covariates are summary characteristics, defined for locations in the observation window, either for the observations' locations or for all locations, reflecting inter-individual spatial behavior such as local interaction or competition. This is different, although is in spirit with, the work of \cite{illian2012toolbox}, \cite{illian2013} and \cite{raeisi2020}, that include local interaction on log-Gaussian Cox processes by including a constructed covariate based on the nearest point distance on the point processes intensity.

Whenever there is preferential sampling, observations are closer in space where the spatial phenomena of interest is higher or smaller, depending on the type of preferentiability, than it would be expected if the sampling was not preferential. This led us to consider in this work, three different constructed covariates to deal with the preferential sampling effect, based on average distances between the locations of each observation and the other observations chosen according to different criteria:

\begin{itemize}
	\item[$\bullet$] \texttt{Construct 1}: averaged distances to $k$ nearest neighbours;
	
	\item[$\bullet$] \texttt{Construct 2}: averaged distances to all neighbours;
	
	\item[$\bullet$] \texttt{Construct 3}: averaged distances to all neighbours that are less than a fixed distance $d$ apart.
\end{itemize}

In \texttt{Construct 1}, for each observation on location $x$ in the observation pattern $X= (x_1, \cdots, x_n)$, we calculate the average distance to the $k$ nearest points in $X$ as
$$\mbox{ \texttt{Const}}_1(x)=\frac{1}{k} \sum_{j \in \partial_x} ||x - x_j ||.$$
where $|| \cdot ||$ denotes the Euclidean distance and $\partial_x$ denotes the set of indices of the $k$ nearest observation points to $x$ in the observation pattern $X$.

\vspace{0.2cm}

In \texttt{Construct 2}, for each observation on location $x$ in the observation pattern $X$, we calculate the average of the distances to all neighbours in the observation pattern $X = (x_1, \cdots, x_n)$ as

$$\mbox{ \texttt{Const}}_2(x)=\frac{1}{n-1} \sum\limits_{\substack{j=1 \\ x_j\neq x}}^n  ||x - x_j ||.$$

\vspace{0.2cm}

In \texttt{Construct 3},  for each observation on location $x$ in the observation pattern $X$, we calculate the average of the distances to all neighbours in the observation pattern $X = (x_1, \cdots, x_n)$ that are less than a fixed distance $d$ apart.

$$\mbox{ \texttt{Const}}_3(x)=\frac{1}{\#(\partial_x^d)} \sum_{j \in \partial_x^d} ||x - x_j ||.$$
where $\partial_x^d$ denotes the set of indices of the observation points in the observation pattern $X$ that are less than distance $d$ apart from $x$ and $\#(\partial_x^d)$ denotes the number elements in that set.
Depending on the chosen value of $d$, there may be observed points $x$ for which there is no neighbour less than a distance $d$. In these situations, we set \texttt{Const}$_3(x)$ equal to $d$.

\vspace{0.2cm}

For datasets subject to preferential sampling, we introduce a practical and accessible methodology aimed at mitigating its adverse effects without resorting to complex joint modelling frameworks. The core idea is to incorporate a constructed covariate, as the ones just described, in the standard geostatistical model that would ordinarily be applied under the assumption of non-preferential sampling, model (\ref{eq_geo}) with covariates.

These constructed covariates are designed to act as proxies for the latent dependence between the sampling process and the spatial variable of interest. By explicitly modelling this dependence through an additional covariate, we can often recover the validity of standard inference procedures. In other words, if the constructed covariate sufficiently captures the structure introduced by preferential sampling, the resulting model can still yield unbiased and interpretable estimates without the need for computationally demanding joint models, which may be difficult to specify, fit, or interpret in practice.

This approach offers several advantages. First, it provides a simple and implementable alternative for researchers and practitioners who may lack access to specialized software or the computational resources required by full joint models. Second, it retains interpretability and transparency, as it builds upon familiar modelling tools already in use within the geostatistical community. Third, it offers flexibility, since different types of constructed covariates can be defined depending on the spatial context, the sampling density, or the suspected mechanism of preferentiality.

To assess whether the inclusion of the constructed covariate has successfully accounted for the effects of preferential sampling, we employ the Means and Locations Correlation (MLC) test, as proposed by \cite{natario2026testingpreferentialsampling}. This test evaluates the residual correlation between the spatial locations and the observed values (marks). If the test fails to detect significant dependence after the constructed covariate has been included, this provides empirical support for proceeding with standard statistical methods.

In the following section, we demonstrate the utility and effectiveness of this methodology through an extensive set of numerical studies, under varying degrees of preferentiality and sampling intensities. These results highlight the conditions under which the proposed approach is most effective, and they offer guidance on its practical application in real-world scenarios.

\section {Means and Locations Correlation test} \label{MLCtest}

\begin{figure}[!htb]
	\begin{center}
		\includegraphics[width=0.5\linewidth]{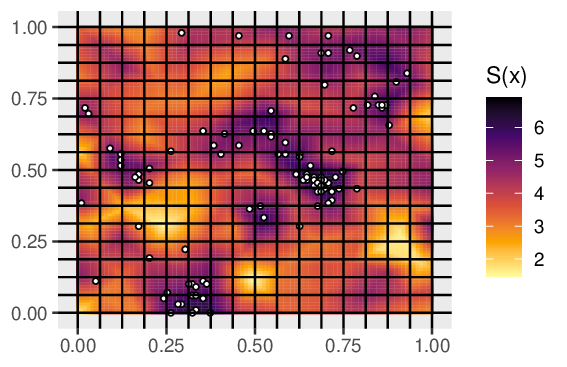}
		\caption{Considering a partition of the observation window,  MLC test performs a correlation test between the number of observed points and the average observed measures in each cell.}
		\label{fig:figgrid}
	\end{center}
\end{figure}

To test the null hypothesis of stochastic independence between the sampling locations and the spatial process, we use the Means and Locations Correlation (MLC) test, proposed by \cite{natario2026testingpreferentialsampling}. This non-parametric test is based on the idea that, if dependence exists between the sampling locations and the spatial process, then, after partitioning the study region into cells, the number of sampled points within each cell is correlated with the mean value of the corresponding observations of the process (see Figure \ref{fig:figgrid}). The MLC test can be implemented under either a frequentist or bayesian framework.

The MLC test performs a Spearman rank correlation test between the number of sampled points in each cell and the average observed values within the same cell, with an adjustment for ties. The use of Spearman’s correlation does not require the observations $Y$ to follow a Gaussian distribution or even to be continuous.

\cite{natario2026testingpreferentialsampling} conducted a large simulation study considering both regular and irregularly shaped regions, different levels of preferentiality, scenarios with and without covariate effects, and different grid sizes. In this work, the grid side length $l$ is defined as the square root of the average area per sampling point, namely $l =\sqrt{\dfrac{Area_D}{n}}$, where $Area_D$ denotes the total area of the study domain, a choice that proved to be a sensible one for this test.

The results reported in \cite{natario2026testingpreferentialsampling}, in its classical approach, were highly encouraging. Preferential sampling was correctly identified in almost 100\% of the simulated datasets generated with $\beta \in \left\lbrace -2,-1.5,-1,1,1.5,2\right\rbrace $. This proportion decreased to approximately 80\% for datasets simulated with $\beta=0.5$ whereas for $\beta=-0.5$ it drops further to about 70\%. When  $\beta = 0$, the test incorrectly rejects the null hypothesis of stochastic independence in 4\% of the simulated replicates. Comparable findings were obtained for the Bayesian formulation of the MLC test, showing performance broadly consistent with that of the classical approach. However, the Bayesian implementation is more computationally demanding. Therefore, given the large number of simulations and replicates considered in the numerical studies presented in the following section, the frequentist version of the MLC test is adopted, for different construct covariates choice. Nevertheless, the Bayesian formulation is subsequently applied to the construct covariate, which exhibits the best performance, and to the real case studies considered in this work.

\section{Numerical Studies} \label{numerical1}

\begin{figure}[!htb]
	\begin{center}
		\includegraphics[width=0.49\linewidth]{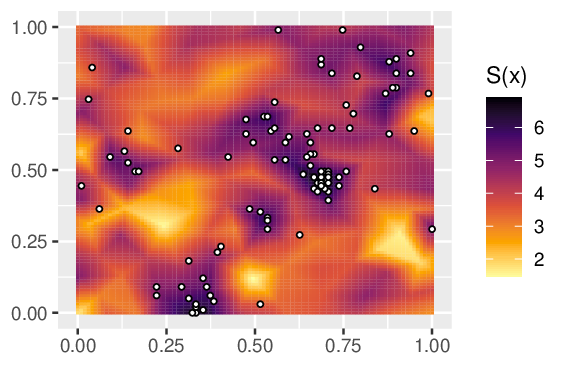}
		\includegraphics[width=0.49\linewidth]{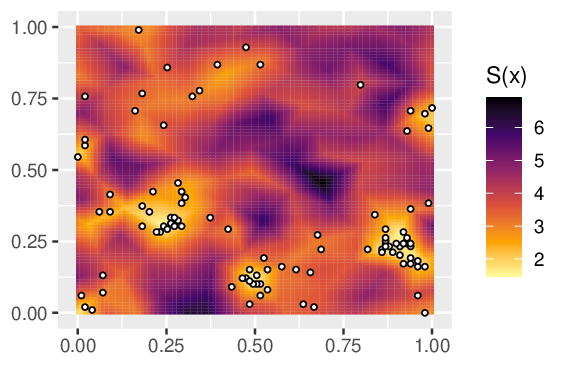}\\
		\includegraphics[width=0.49\linewidth]{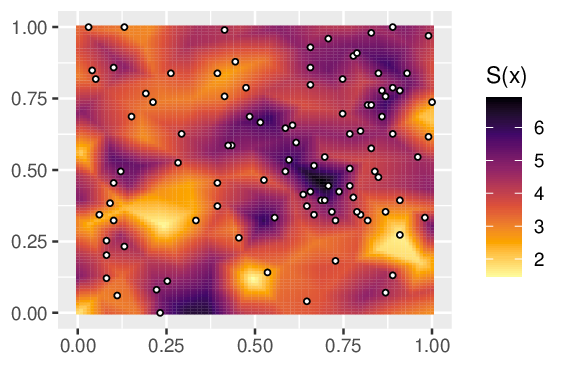}
		\caption{Simulated sampled data sets from model (\ref{eq_lambda}), assuming $\beta=2$ (top left); $\beta=-2$ (top right); $\beta=0.5$ (bottom).}
		
		\label{fig:beta}
	\end{center}
\end{figure}

This Section documents the performance of the proposed methodology of including a constructed covariate to eliminate the effect of preferential sampling, across a range of simulated data settings. We consider eight quite different degrees of preferentiality, corresponding to simulated data from model (\ref{eq_lambda}) with $\beta \in \left\lbrace -4, -2, -1, -0.5, 0.5, 1, 2, 4 \right\rbrace $, and three different data sample sizes, namely $n=50$, $n=100$ and $n=250$. Each experimental setting is repeated 100 times. To illustrate these sampling schemes, we represent in Figure \ref{fig:beta}, three different sets of 100 sampled points, simulated  for $\beta$ equal to 2, -2 and 0.5, illustrating quite well the preferential sampling problem. The code used can be find in \url{https://github.com/Andreia-Monteiro/Accounting-PS}. \\

\vspace{-0.3cm}

\subsection {\texttt{Construct 1}: simulation on a regularly shaped domain} \label{Const1reg}

The first simulation study aims to investigate if \texttt{Construct 1}, given for each observation as the averaged distances to its $k$ nearest neighbours, accounts for preferential sampling whenever that is the case. A regularly shaped domain of interest is considered, $\left\lbrace S(x) : x \in \mathbb{R}^2\right\rbrace $ is taken as a realization of a Gaussian process with mean $\mu=4$, variance $\sigma^2 =1.5$ and Matérn correlation with scale parameter $\phi=0.15$ and shape parameter $\kappa=1$. The sampled locations are generated from a log Gaussian Cox process with intensity $\exp(\beta S(x))$ and nugget variance $\tau^2=0.1$. In the study, all combinations of the following parameters are evaluated:

\begin{itemize}
	\item Sample size $n \in \left\lbrace 50, 100, 250\right\rbrace $;
	
	\item Degree of preferentiality $\beta \in \left\lbrace -4, -2, -1, -0.5, 0.5, 1, 2, 4 \right\rbrace $;
	
	\item Number of neighbours considered in the definition of \texttt{Construct 1}, $k \in \big \{ [n \times 5\%], [n \times 10\%] \big \} $, where $[\cdot]$ represents the integer part.
\end{itemize}

\begin{figure}[!htb]
	\centering
	\includegraphics[width=0.5\linewidth]{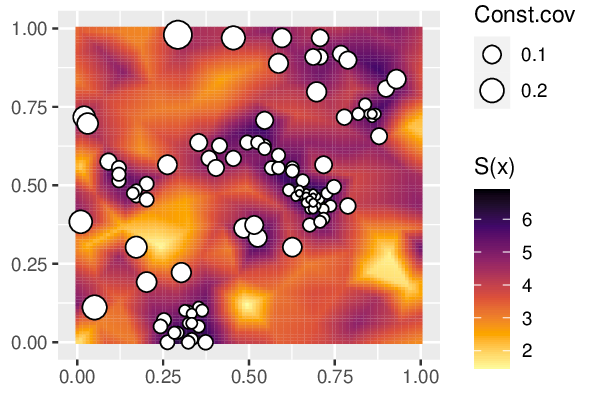}
	\caption{Simulated pattern with $\beta=2$ and $n=100$ and the associated constructed covariate \texttt{Construct 1} for this pattern.}
	\label{fig:rplotconstruct}
\end{figure}

Figure \ref{fig:rplotconstruct} illustrates a simulated pattern obtained considering $\beta=2$ and $n=100$ and the associated constructed covariate \texttt{Construct 1} for this pattern.

A total of 100 independent samples were simulated for the different degrees of preferentiality and sample sizes. Table \ref{tab:table0} summarizes the percentage of replicas in which preferential sampling was detected with MLC test, considering $5\%$ significance level.

\begin{table}[!htb]
	\begin{center}
		\caption{Percentage of replicas in which preferential sampling was detected with MLC test (without including contructed covariate)
			\label{tab:table0}}
		\begin{tabular}{lccc}
			\hline\noalign{\smallskip}
			& $n=50$ & $n=100$ & $n=250$ \\
			\hline \noalign{\smallskip}
			$\beta=-4$  & 43\%  & 90\%  & 100\% \\
			$\beta=-2$  & 60\%  & 99\%  & 100\% \\
			$\beta=-1$  & 52\%  & 95\%  & 100\% \\
			$\beta=-0.5$  & 22\%  & 52\%  & 96\% \\
			$\beta=0.5$  & 20\%  & 41\%  & 95\% \\
			$\beta=1$  & 47\%  & 92\% & 100\%\\
			$\beta=2$  & 69\%  & 100\% & 100\% \\
			$\beta=4$  & 44\%  & 93\%  & 100\% \\
			\noalign{\smallskip}\hline
		\end{tabular}
	\end{center}
\end{table}

Considering \texttt{Construct 1} given by the average distance of the 5\% of the nearest neighbour observations, we applied for each simulated data set the MLC test to the residuals of the geostatistical model (\ref{eq_geo}) fitted with the constructed covariate, under a Bayesian framework as described in \cite{krainski2018}. For the spatial parameters, the Penalised Complexity Priors (PC Priors) introduced by \cite{simpson2017}  were used, according with some aspects, such as the histogram and the median of the distance between the coordinates of the observations and the practical range obtained in
the Matérn variogram. In respect to the PC prior for the standard deviation, $P[\sigma > 1] = 0.01$ was applied. Therefore, for the range $\left( r=\frac{\sqrt{8}}{\kappa}\right) $, the PC prior $P[r < 0.05] = 0.01$ was considered. For the remaining parameters, we used the default priors of R-INLA, the prior on the intercept is a uniform distribution, the prior on the coefficients is a Gaussian and the prior on the precision is a Gamma.

Table \ref{tab:table1} summarizes the percentage of replicas in which the inclusion of the constructed covariate mitigate the effect of preferential sampling, considering $5\%$ significance level. By analysing Table \ref{tab:table1}, the results are quite satisfactory and we believe that the inclusion of the constructed covariate is able to explain the stochastic dependence of sampling locations on the spatial variable under study.


\begin{table}[!htb]
	\begin{center}
		\caption{Percentage of replicas in which the inclusion of the constructed covariate \texttt{Const}$_1$ mitigate the effect of preferential sampling.
			\label{tab:table1}}
		\begin{tabular}{lccc}
			\hline\noalign{\smallskip}
			
			& $n=50$ & $n=100$ & $n=250$ \\
			\hline \noalign{\smallskip}
			$\beta=-4$  & 97\%  & 93\%  & 84\% \\
			$\beta=-2$  & 98\%  & 94\%  & 85\% \\
			$\beta=-1$  & 98\%  & 99\%  & 96\% \\
			$\beta=-0.5$  & 99\%  & 98\%  & 98\% \\
			$\beta=0.5$  & 97\%  & 96\%  & 99\% \\
			$\beta=1$  &97\%  & 99\%  & 94\% \\
			$\beta=2$  &97\%  & 95\% & 97\%\\
			$\beta=4$  & 98\%  & 95\% & 95\% \\
			
			\noalign{\smallskip}\hline
		\end{tabular}
	\end{center}
\end{table}

Considering the average distance to 10\% of nearest neighbours for the constructed covariate, the results were very similar to those presented here with 5\%.

\subsection {\texttt{Construct 2}: simulation on a regularly shaped domain}

In this simulation study we consider the same simulated data as described in Section \ref{Const1reg}.

Taking for each observed point the average distance to all $(n-1)$ neighbours we define the constructed covariate. Considering the 100 independent simulated samples, for the different degrees of preferentiality and sample sizes, we applied the MLC test to the residuals of the geostatistical model (\ref{eq_geo}) fitted with the constructed covariate for each of these data sets, under a Bayesian framework. Table \ref{tab:table1Cov2} summarizes the percentage of replicas in which the inclusion of the constructed covariate mitigate the effect of preferential sampling, considering $5\%$ significance level. By analysing Table \ref{tab:table1Cov2}, the results are worse than those obtained with the constructed covariate \texttt{Const}$_1$, namely in cases with a higher degree of preferentiability.

In order to avoid the influence of extreme distances we also considered the median distance to $(n-1)$ neighbours. However, the results were quite similar.


\begin{table}[!htb]
	\begin{center}
		\caption{Percentage of replicas in which the inclusion of the constructed covariate  \texttt{Const}$_2$ mitigate the effect of preferential sampling. 
			\label{tab:table1Cov2}}
		\begin{tabular}{lccc}
			\hline\noalign{\smallskip}
			
			& $n=50$ & $n=100$ & $n=250$ \\
			\hline \noalign{\smallskip}
			$\beta=-4$  & 88\%  & 76\%  & 82\% \\
			$\beta=-2$  & 77\%  & 75\%  & 73\% \\
			$\beta=-1$  & 86\%  & 79\%  & 83\% \\
			$\beta=-0.5$  & 92\%  & 85\%  & 85\% \\
			$\beta=0.5$  & 95\%  & 90\%  & 93\% \\
			$\beta=1$  & 83\%  & 67\%  & 73\% \\
			$\beta=2$  & 82\%  & 68\% & 71\%\\
			$\beta=4$  & 91\%  & 86\% & 79\% \\
			
			\noalign{\smallskip}\hline
		\end{tabular}
	\end{center}
\end{table}

\subsection {\texttt{Construct 3}: simulation on a regularly shaped domain}

In this simulation study we consider the same simulated data as described in Section \ref{Const1reg}.

For each observed point $x$, we calculate average of the distances to all neighbours, in the observation pattern $X = (x_1, \cdots, x_n)$, that are closer than a distance $d$ from $x$. Several values for $d$ are considered, the best results are those presented for $d = 0.05$ (5\% on the regularly shaped domain side).

Considering the 100 independent simulated samples, for the different degrees of preferentiality and sample sizes, we applied the MLC test to the residuals of the geostatistical model (\ref{eq_geo}) fitted with the constructed covariate for each of these data sets, under a Bayesian framework. Table \ref{tab:table1Cov3} summarizes the percentage of replicas in which the inclusion of the constructed covariate mitigate the effect of preferential sampling, considering $5\%$ significance level. By analysing Table \ref{tab:table1Cov3}, the results are worse than those obtained with the constructed covariate \texttt{Const}$_1$, but better than those obtained with the constructed covariate \texttt{Const}$_2$.


\begin{table}[!htb]
	\begin{center}
		\caption{Percentage of replicas in which the inclusion of the constructed covariate  \texttt{Const}$_3$ mitigate the effect of preferential sampling. 
			\label{tab:table1Cov3}}
		\begin{tabular}{lccc}
			\hline\noalign{\smallskip}
			
			& $n=50$ & $n=100$ & $n=250$ \\
			\hline \noalign{\smallskip}
			$\beta=-4$  & 89\%  & 93\%  & 92\% \\
			$\beta=-2$  & 91\%  & 75\%  & 85\% \\
			$\beta=-1$  & 86\%  & 89\%  & 90\% \\
			$\beta=-0.5$  & 94\%  & 88\%  & 98\% \\
			$\beta=0.5$  & 95\%  & 94\%  & 93\% \\
			$\beta=1$  & 90\%  & 85\%  & 90\% \\
			$\beta=2$  & 91\%  & 81\% & 91\%\\
			$\beta=4$  & 94\%  & 87\% & 93\% \\
			
			\noalign{\smallskip}\hline
		\end{tabular}
	\end{center}
\end{table}

As \texttt{Const}$_1$ was the one with the best results, the subsequent analyses continue with the use of this constructed covariate.

\subsection {\texttt{Construct 1}: simulation on an irregularly shaped domain}

\begin{figure}[!htb]
	\begin{center}
		\includegraphics[width=0.49\linewidth]{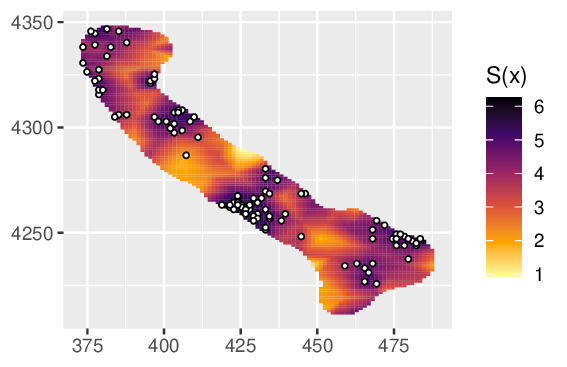}
		\includegraphics[width=0.49\linewidth]{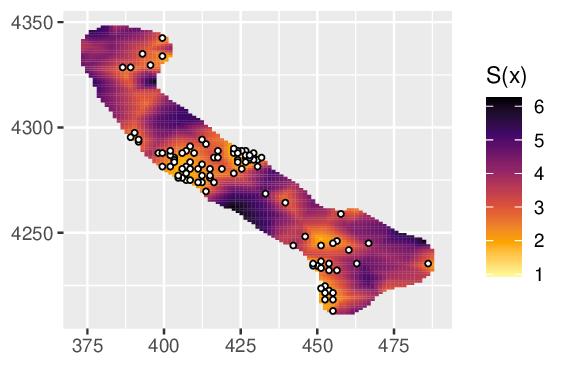}\\
		\includegraphics[width=0.49\linewidth]{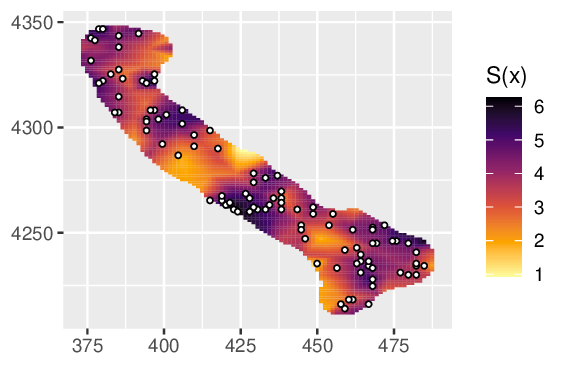}
		\caption{Simulated sampled sets assuming $\beta=2$ (top left); $\beta=-2$ (top right); $\beta=0.5$ (bottom).
			\label{fig:fish_area}}
	\end{center}
\end{figure}

The second simulation study presented is based on a real data set of captures of Black Scabbardfish off the Portuguese coast, a deep-water fish species that lives at depths greater than 800m, therefore has its captures confined into a very irregular shaped region.  On the portuguese coast, Black Scabbardfish constitutes an important commercial resource. In the absence of dedicated deep-water research surveys in this area, the spatial distribution of its abundance is mainly inferred from commercial deep-water longline fishery operating along the continental slope, \cite{paula2023}.
For this region,
in the simulation study, 100 different spatial patterns were generated from a Gaussian process with mean $\mu=4$, variance $\sigma^2 =1.5$ and Matérn correlation with scale parameter $\phi=15$ and shape parameter $\kappa=1$. The sampled locations were simulated accordingly as before.To illustrate these situation, we represent in Figure \ref{fig:fish_area}, 100 sampled points, for $\beta$ equals 2, -2 and 0.5. Table \ref{tab:table5} summarizes the percentage of replicas in which preferential sampling was detected with MLC test, considering $5\%$ significance level.

\begin{table}[!htb]
	\begin{center}
		\caption{Percentage of replicas in which preferential sampling was detected with MLC test (without including constructed covariate)
			\label{tab:table5}}
		
		\begin{tabular}{lcccc}
			\hline \noalign{\smallskip}
			
			& $n=50$ & $n=100$ & $n=250$ &  $n=733$\\
			\hline \noalign{\smallskip}
			$\beta=2$  & 70\%  & 99\%  & 100\% & 100\%\\
			$\beta=-2$  & 76\%  & 97\% & 100\% & 100\%\\
			$\beta=0.5$  & 76\%  & 60\% & 88\% & 100\%\\
			
			\noalign{\smallskip}\hline
		\end{tabular}
		
	\end{center}
\end{table}

Considering the average distance to 5\% of the nearest neighbours for the constructed covariate and simulating for the different degrees of preferentiality and sample sizes, a total of 100 independent samples, Table \ref{tab:table2} summarizes the percentage of replicas in which the inclusion of the constructed covariate mitigate the effect of preferential sampling, considering $5\%$ significance level.

\begin{table}[!htb]
	\begin{center}
		\caption{Percentage of replicas in which the inclusion of the constructed covariate \texttt{Const}$_1$ mitigate the effect of preferential sampling.
			\label{tab:table2}}
		
		\begin{tabular}{lcccc}
			\hline \noalign{\smallskip}
			
			& $n=50$ & $n=100$ & $n=250$ &  $n=733$\\
			\hline \noalign{\smallskip}
			$\beta=2$  & 99\%  & 97\%  & 96\% & 98\%\\
			$\beta=-2$  & 99\%  & 96\% & 92\% & 99\%\\
			$\beta=0.5$  & 100\%  & 98\% & 98\% & 97\%\\
			
			\noalign{\smallskip}\hline
		\end{tabular}
		
	\end{center}
\end{table}

By analysing Table \ref{tab:table2}, once again, the results are quite satisfactory and the inclusion of the constructed covariate is able to explain the stochastic dependence
of sampling locations on the spatial variable under study.

\subsection {Model comparison}

In order to illustrate the effectiveness of including the constructed covariate \texttt{Const}$_1$ in the model in terms of the predictive power, we consider a model comparison simulation study. 

For each of the 100 simulated data sets on regularly shaped domain we fitted the conventional geostatistical model with the inclusion of the constructed covariate \texttt{Const}$_1$ (MC) and  we fitted  the joint model proposed by \cite{diggle2010} using a Bayesian approach (MP).

To compare their performance we used  the log‐conditional predictive ordinates (LCPO), \cite{roos2011}, LCPO is a “leave-one-out” cross‐validation index to assess the predictive power of the model. Lower values of LCPO suggest better model performance, \cite{pennino2018}.

\begin{figure}[!htb]
	\centering
	
	\includegraphics[width=0.49\linewidth]{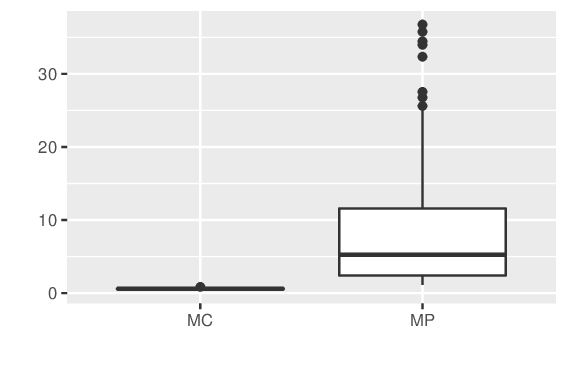}
	\includegraphics[width=0.49\linewidth]{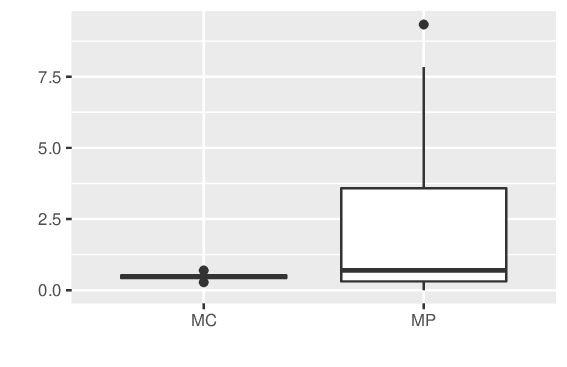}
	\caption{LCPO scores for MC and MP, $\beta=2$, $n = 100 $ (left) and $n = 200$ (right)}
	\label{fig:comparen100}
\end{figure}

\begin{figure}[!htb]
	\centering
	\includegraphics[width=0.49\linewidth]{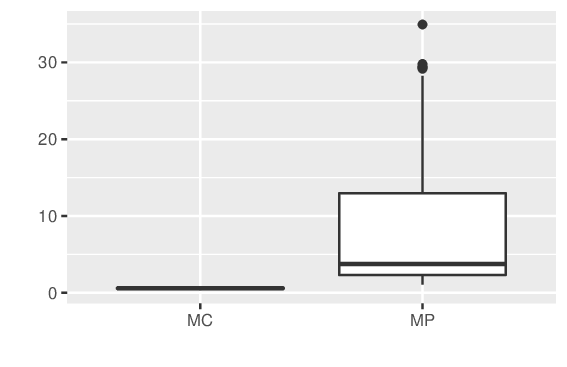}
	\includegraphics[width=0.49\linewidth]{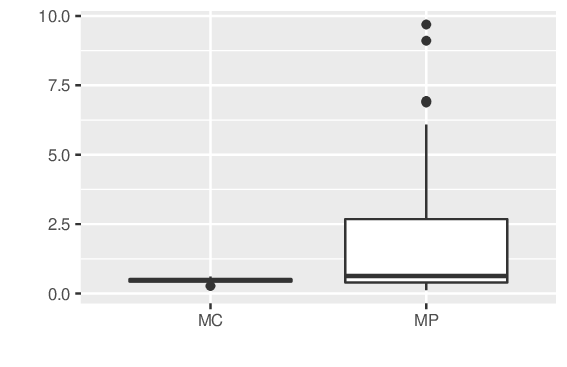}
	\caption{LCPO scores for MC and MP, $\beta=-2$, $n = 100 $ (left) and $n = 200$ (right)}
	\label{fig:comparen250}
\end{figure}

The simulation study to compare models MC and MP was carried out for $\beta \in \left\lbrace -2, 2\right\rbrace $ and with a sample size $n \in \left\lbrace 100, 200\right\rbrace $, MP model presented some problems in the adjustment, possibly due to the too regular data simulation process and we reduced the sample size to $n=200$. Figure \ref{fig:comparen100} shows LCPO scores for MC and MP models, with $\beta = 2$ for $n \in \left\lbrace 100, 200\right\rbrace $. Figure \ref{fig:comparen250} shows LCPO scores with $\beta = -2$ for $n \in \left\lbrace 100, 200\right\rbrace $. From these Figures, it can be appreciated a fit improvement of MC over MP in terms of the predictive power measured by LCPO.

As far as estimation is concerned both models (MC and MP) return good estimates, although MP model depicts estimates slightly closer to the true values of the parameters.

\section {A Completely Bayesian Formulation}


In this section, we apply the Bayesian version of the MLC test, as proposed by \cite{natario2026testingpreferentialsampling}, by making use of Bayes factors. Specifically, they follow the rank-based hypothesis testing approach for Spearman’s rank correlation coefficient, $\rho_S$, developed by \cite{van2020bayesian}. This method relies on Bayes factors to assess competing hypotheses about the presence or absence of correlation in the data. A central innovation in their framework is the use of latent normal models, which enables the transformation of ordinal or rank-based data into a parametric setting. By embedding the rank correlation problem within a latent variable framework, Bayesian inference becomes tractable and allows for a principled comparison of models representing independence versus association.

The Bayesian approach not only complements the frequentist test presented earlier but also provides additional insights into the strength of evidence for preferential sampling. While the frequentist MLC test is based on rejecting a null hypothesis at a predetermined significance level, the Bayesian framework allows us to quantify how strongly the observed data support one hypothesis relative to the other. This is particularly useful in simulation studies, where the Bayes factor can highlight subtle but systematic deviations from randomness that may not be evident in conventional significance testing.

The results of the Bayesian test, considering a regularly shaped domain and the inclusion of the constructed covariate \texttt{Const}$_1$, are summarized in Table \ref{tab:table7}, where they are directly compared with those obtained under the frequentist approach. For $\beta = 4$ and $\beta = -4$, with $n=50$, some simulations produced \texttt{NA} values due to convergence and mixing difficulties in the Monte Carlo Markov Chain (MCMC) algorithm used to perform the test, particularly as a result of the high autocorrelation observed in the generated chains. These issues are likely related to the strong preferentiality scenarios combined with the relatively small sample sizes.

Overall, both approaches lead to similar conclusions regarding the presence or absence of preferential sampling. However, the Bayesian implementation is considerably more computationally demanding, reflecting the additional burden associated with latent-variable modeling and Bayes factor estimation.


\begin{table}[!htb]
	\begin{center}
		\caption{Percentage of replicas in which the inclusion of the constructed covariate \texttt{Const}$_1$ mitigate the effect of preferential sampling.
			\label{tab:table7}}
		\begin{tabular}{lccc}
			\hline\noalign{\smallskip}
			& $n=50$ & $n=100$ & $n=250$ \\
			\hline \noalign{\smallskip}
			$\beta=-4$  & NA  & 87\%  & 94\% \\
			$\beta=-2$  & 78\%  & 92\%  & 96\% \\
			$\beta=-1$  & 95\%  & 96\%  & 98\% \\
			$\beta=-0.5$  & 99\%  & 98\%  & 98\% \\
			$\beta=0.5$  & 96\%  & 97\%  & 99\% \\
			$\beta=1$  & 95\%  & 93\% & 96\%\\
			$\beta=2$  & 96\%  & 95\% & 95\% \\
			$\beta=4$  & NA  & 90\%  & 89\% \\
			\noalign{\smallskip}\hline
		\end{tabular}
	\end{center}
\end{table}


\section {Case Studies}

The effectiveness of the proposed methodology is further illustrated through its application to two real-world case studies, each exemplifying a different form of preferential sampling. These contrasting examples provide a robust test of the flexibility and practical value of our approach.

In the first case study, we consider fishery catch data, a classic example of positive preferential sampling. Here, sampling locations i.e., fishing efforts, are not randomly distributed across the study area but are instead consciously selected by fishermen based on their expectations of high fish abundance. As a result, the observed data are more frequently recorded in regions with presumed higher values of the variable of interest, reflecting a clear dependence between location choice and the underlying spatial process. This setting offers a natural and meaningful context in which to apply our constructed covariate approach, as the dependence between sampling locations and observed values is directly driven by human decision-making.

The second case study comprises measurements of heavy metal concentrations in moss samples. Since mosses accumulate atmospheric pollutants, they serve as biomonitors of airborne heavy metal deposition. A key feature of this dataset is the presence of negative preferential sampling, with higher sampling intensity in areas expected to have lower concentrations. This type of bias, though less intuitive, has been documented in previous studies, including \cite{watson2020}, and may arise in environmental monitoring contexts where regions perceived as ``problematic" or at higher risk are oversampled at the expense of others. This imbalance can distort the overall picture of spatial variability and lead to misleading conclusions if not properly addressed.

By applying our methodology to both scenarios, we aim to demonstrate its versatility and robustness across different sampling behaviours, highlighting how the inclusion of constructed covariates can adjust for both positive and negative forms of preferentiality. These applications also underscore the practical advantages of our approach: retaining model simplicity, reducing computational demands, and enabling reliable inference even in the presence of complex sampling designs.

\subsection{Black Scabbardfish catches}

\begin{figure}[!h]
	
	\centering
	\includegraphics[width=0.6\linewidth]{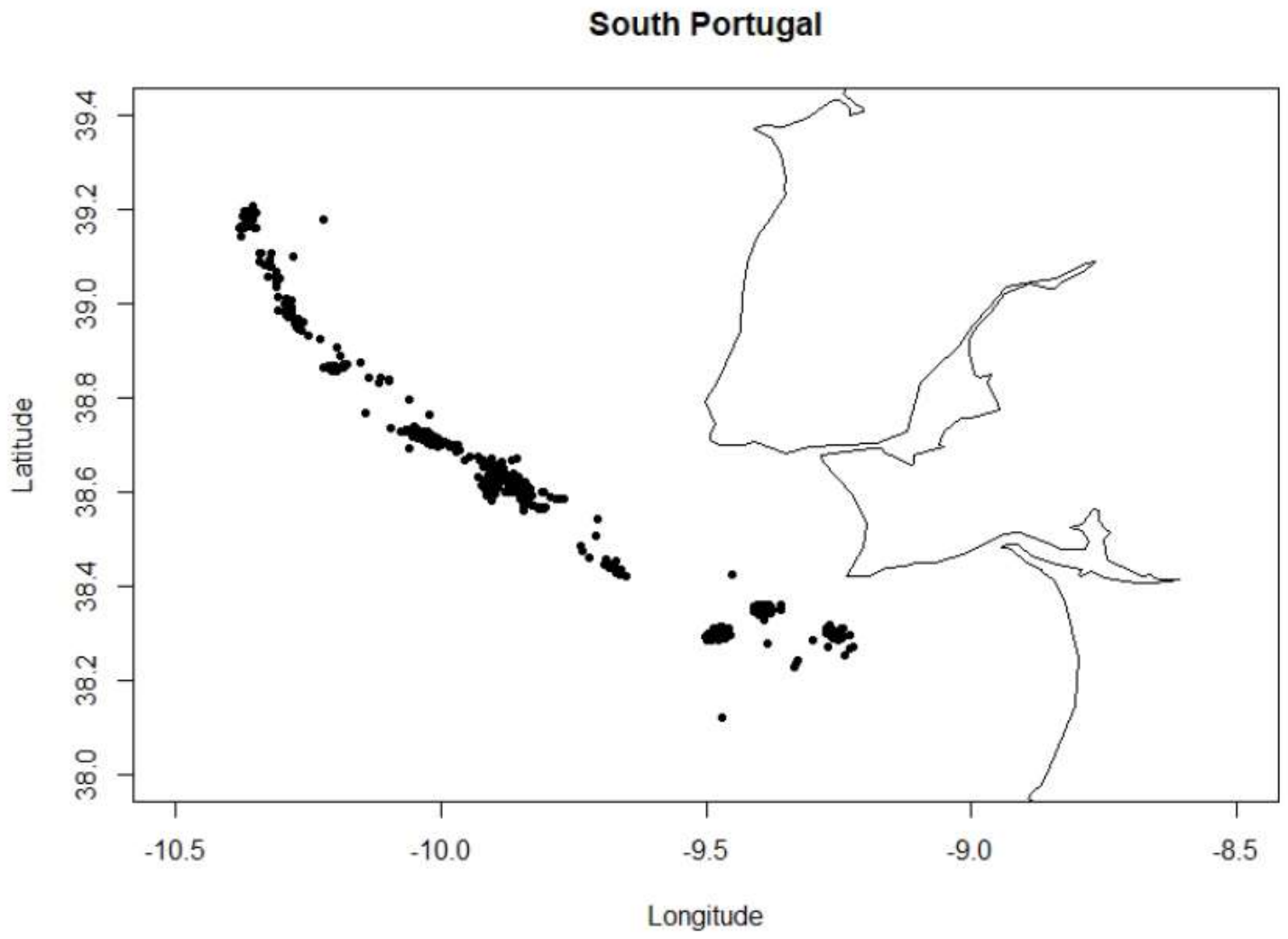}
	\caption{Locations of the BSF catches off the Portuguese coast.}
	\label{fig:fig1}
\end{figure}

We illustrate the proposed methodology using a real dataset provided by the Instituto Português do Mar e da Atmosfera (IPMA) and consists of fishery-dependent observations collected by the Portuguese commercial longline fleet operating along the Portuguese continental slope. Following \cite{andre2020}, we consider a subset of the original dataset comprising fishing hauls carried out south of latitude ($39.3^\circ$) N between September and February during the period 2009-2013, yielding a total of 733 observations.  Data include the Black Scabbardfish (BSF) catches (in kg) by fishing haul of the longline fishing fleet but also include the location of each fishing haul, Figure \ref{fig:fig1}. As fishing vessels intentionally select fishing grounds where higher catches are expected, the sampling locations are not independent of the underlying abundance process, making this dataset a representative example of positive preferential sampling.

The BSF captures in Portuguese waters were previously modelled, taking the sampling preferentiality into account, using a Bayesian approach and INLA methodology, considereing stochastic partial differential equations for geostatistical data, jointly with a Log-Cox point process model, \cite{paula2022}.

For the Black Scabbardfish dataset, the classical version of the MLC test yielded a p-value of 0.0550, placing the result at the boundary of the 5\% significance threshold. This indicates only weak evidence of preferential sampling and does not justify rejecting the null hypothesis of stochastic independence between sampling locations and observed values. After including the constructed covariate \texttt{Const}$_1$ (average distance to the 5\% nearest neighbours) in the geostatistical model, the p‑value increased to 0.1536, showing that the residual dependence between locations and marks is no longer statistically significant. This supports the adequacy of the constructed covariate in mitigating the preferential sampling effect.

We also applied the Bayesian MLC test to this dataset. Without the constructed covariate, the Bayes factor was 0.5077, which does not provide evidence in favour of preferential sampling. This behaviour is consistent with the low underlying degree of preferentiality expected in this case ($\beta = 0.5$). Indeed, in \cite{andre2020PS}, preferential sampling with an estimated degree of $\beta = 0.5$ was detected using the joint models (\ref{eq_geo}) and (\ref{eq_lambda}), reinforcing the interpretation that the BSF data exhibit only mild preferentiality.
As reported in \cite{natario2026testingpreferentialsampling}, when $\beta = 0.5$ the classical version of the MLC test detects preferential sampling in approximately 80\% of simulated replicates, whereas the Bayesian version detects it in only 70\%, reflecting its more conservative nature.

When the constructed covariate was included, the Bayes factor further decreased to 0.2028, reinforcing the conclusion that no residual dependence remains between sampling intensity and observed values. This result is fully aligned with the classical analysis and provides additional Bayesian evidence that the constructed covariate successfully accounts for the preferential sampling mechanism.

Overall, both the classical and Bayesian analyses indicate that the inclusion of the constructed covariate effectively removes the dependence between sampling locations and the spatial process, thereby supporting the use of standard geostatistical modelling for this dataset.

\subsection{Galicia lead concentrations}

The second dataset, available from the discontinued R-package PrevMap, consists of the concentrations of lead in moss samples collected in Galicia, northern Spain, in 1997. The concentration is measured in micrograms per gram of dry moss, \cite{watson2021}. The 1997 locations were previously shown to have been preferentially sampled, \cite{diggle2010} and \cite{watson2021}.

\begin{figure}[!h]
	
	\centering
	\includegraphics[width=0.5\linewidth]{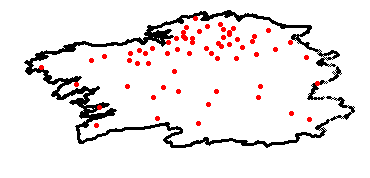}
	\caption{Sampled locations of lead concentrations in Galicia, northern Spain.}
	\label{fig:fig2}
\end{figure}

Figure \ref{fig:fig2} shows the sampling locations, where a higher density of observations is evident in northern Galicia, an area with lower concentrations of lead.

For the Galicia lead dataset, the classical version of the MLC test yields a p‑value of 0.0593, which lies close to the conventional 5\% significance threshold. This result suggests a tendency towards preferential sampling, although the evidence is not strong enough to draw a definitive conclusion.

When the constructed covariate \texttt{Const}$_1$ is included in the geostatistical model, the classical MLC test clearly fails to reject the null hypothesis of independence between sampling locations and the spatial process (p‑value = 0.1911). This indicates that the constructed covariate successfully captures the local structure of the sampling pattern and mitigates the preferential sampling effect.

To complement the frequentist analysis, we also applied the Bayesian MLC test. Without the constructed covariate, the Bayes factor was 1.1238, providing only weak evidence in favour of preferential sampling. After including the constructed covariate, the Bayes factor dropped to 0.4411, indicating that the data no longer support the presence of dependence between sampling intensity and observed lead concentrations. This Bayesian result is fully consistent with the classical analysis and reinforces the conclusion that the constructed covariate effectively accounts for the preferential sampling mechanism.

\section{Final Remarks and Future Work}

The primary goal of this work is to contribute to the growing body of methodologies for addressing preferential sampling, while deliberately avoiding the complexity associated with joint modelling approaches. By proposing a strategy that relies on standard statistical tools, we aim to provide a more accessible and broadly applicable alternative, particularly suited to fields such as biology, ecology, and environmental science, where advanced computational techniques may be difficult to implement or interpret.

Our methodology addresses situations where the sampling design depends on the spatial variable of interest, by incorporating a constructed covariate based on the average distance to the $k$ nearest neighbours. Once this covariate is included in the standard geostatistical model, the analysis can proceed using conventional methods. Extensive simulation studies showed that the approach performs well across a range of conditions, including different sample sizes, varying degrees of preferentiality, and both regular and irregular spatial domains. When compared to the joint model alternative, our proposed method yielded improvements in key aspects, particularly in predictive accuracy while maintaining model simplicity.

Despite these promising results, some challenges remain. One important aspect for further refinement is the optimal selection of the number of nearest neighbours used in the construction of the covariate. Additionally, future work should investigate the extension of this methodology to a wider range of response variables commonly encountered in practical applications. The interpretation of the model may also require careful consideration, as the constructed covariate captures dependence between sampling locations and observed values, rather than representing a directly observed environmental or biological factor.

Future research will focus on addressing these open questions. In particular, we intend to explore more general and adaptive covariates, such as those based on local intensity estimation, which may offer greater flexibility and reduce sensitivity to design choices like the number of neighbours. Nevertheless, despite its limitations, the key strength of the proposed approach lies in its simplicity and accessibility, offering a valuable and practical tool for researchers who need to correct for preferential sampling without resorting to complex models.

\section*{Acknowledgments}

This work is funded by national funds through the FCT - Fundação para a Ciência e a Tecnologia, I.P., under the scope of
the projects UID/00297/2025 (\url{https://doi.org/10.54499/UID/00297/2025}),  UID/PRR/00297/2025 (\url{https:
//doi.org/10.54499/UID/PRR/00297/2025}) (Center for Mathematics and Applications) and through the Project UID/00013/2025 (\url{https://doi.org/10.54499/UID/00013/2025}) (Centre of Mathematics of the University of Minho - CMAT)

The authors would like to express their sincere gratitude to Prof. M. Lucília Carvalho for her invaluable guidance, continuous support, and insightful contributions throughout the development of the methodology presented in this work.

\section* {Data Availability and Code}

Due to the sensitive nature of the Black Scabbardfish catches data, we cannot make them publicly available but researchers interested in using our data may contact the first author of this paper for an individual data use agreement. All code is available through the following repository:
\url{https://github.com/Andreia-Monteiro/Accounting-PS}

\bibliographystyle{unsrtnat}
\bibliography{references1}

@InProceedings{andre2020,
  author = {Andr{\'e}, L{\'i}dia Maria and
            Figueiredo, Ivone and
            Carvalho, M. Luc{\'i}lia and
            Sim{\~o}es, Paula and
            Nat{\'a}rio, Isabel},
  editor = {Gervasi, Osvaldo and
            Murgante, Beniamino and
            Misra, Sanjay and
            Garau, Chiara and
            Ble{\v{c}}i{\'c}, Ivan and
            Taniar, David and
            Apduhan, Bernady O. and
            Rocha, Ana Maria A. C. and
            Tarantino, Eufemia and
            Torre, Carmelo Maria and
            Karaca, Yeliz},
  title = {Spatial Modelling of Black Scabbardfish Fishery Off the Portuguese Coast},
  booktitle = {Computational Science and Its Applications -- ICCSA 2020},
  publisher = {Springer International Publishing},
  address = {Cham},
  year = {2020},
  pages = {332--344},
  doi = {10.1007/978-3-030-58799-4_25}
}

@techreport{andre2020PS,
  author      = {Andr{\'e}, L{\'i}dia Maria and Figueiredo, Ivone and Carvalho, M. Luc{\'i}lia and Sim{\~o}es, Paula and Nat{\'a}rio, Isabel},
  title       = {Spatial Modelling of Black Scabbardfish Fishery under Preferential Sampling of the Portuguese Coast},
  institution = {PREFERENTIAL},
  type        = {Technical Report},
  year        = {2020},
}

@article{conn2017,
  title={Confronting preferential sampling when analysing population distributions: diagnosis and model-based triage},
  author={Conn, Paul B and Thorson, James T and Johnson, Devin S},
  journal={Methods in Ecology and Evolution},
  volume={8},
  number={11},
  pages={1535--1546},
  year={2017},
  publisher={Wiley Online Library}
}

@article{diggle1998,
    author = {Diggle, P. J. and Tawn, J. A. and Moyeed, R. A.},
    title = {Model-Based Geostatistics},
    journal = {Journal of the Royal Statistical Society Series C: Applied Statistics},
    volume = {47},
    number = {3},
    pages = {299-350},
    year = {1998},
    month = {09},
    issn = {0035-9254},
    doi = {10.1111/1467-9876.00113},
    url = {https://doi.org/10.1111/1467-9876.00113},
    eprint = {https://academic.oup.com/jrsssc/article-pdf/47/3/299/48750383/jrsssc_47_3_299.pdf},
}

@book{diggle2007,
  title={Model-based Geostatistics},
  author={Diggle, Peter J and Ribeiro, Paulo J },
  year={2007},
  publisher={Springer}
}

@article{diggle2010,
  title={Geostatistical inference under preferential sampling},
  author={Diggle, Peter J and Menezes, Raquel and Su, Ting-li},
  journal={Journal of the Royal Statistical Society: Series C (Applied Statistics)},
  volume={59},
  number={2},
  pages={191--232},
  year={2010},
doi = {10.1111/j.1467-9876.2009.00701.x},
  url = {https://doi.org/10.1111/j.1467-9876.2009.00701.x},
  publisher={Wiley Online Library}
}

@incollection{diggle2017,
  author ={Diggle, Peter and Giorgi, Emanuele},
  title       = {Preferential sampling of exposure levels},
   booktitle   = {Handbook of Environmental and Ecological Statistics},
  publisher   = {CRC Press},
  year        ={ 2017},
   chapter     = {21},
}

@article{dinsdale2019,
  title={Modelling ocean temperatures from bio-probes under preferential sampling},
  author={Dinsdale, Daniel and Salibian-Barrera, Matias},
  journal={The Annals of Applied Statistics},
  volume={13},
  number={2},
  pages={713--745},
  year={2019},
  publisher={Institute of Mathematical Statistics}
}

@article{frenchetal2021,
title = {Comparing five methods for quantifying abundance and diversity of fish assemblages in seagrass habitat},
journal = {Ecological Indicators},
volume = {124},
pages = {107415},
year = {2021},
author = {Ben French and Shaun Wilson and Thomas Holmes and Alan Kendrick and Michael Rule and Nicole Ryan},
}

@article{gelfand2012,
  title={On the effect of preferential sampling in spatial prediction},
  author={Gelfand, Alan E and Sahu, Sujit K and Holland, David M},
  journal={Environmetrics},
  volume={23},
  number={7},
  pages={565--578},
  year={2012},
  publisher={Wiley Online Library}
}

@article{illian2012toolbox,
  title={A toolbox for fitting complex spatial point process models using integrated nested Laplace approximation (INLA)},
  author={Illian, Janine B and S{\o}rbye, Sigrunn H and Rue, H{\aa}vard},
  journal={The Annals of Applied Statistics},
  pages={1499--1530},
doi={https://doi.org/10.1214/11-AOAS530},
  year={2012},
  publisher={JSTOR}
}

@article{illian2013,
  title={Fitting complex ecological point process models with integrated nested Laplace approximation},
  author={Illian, Janine B and Martino, Sara and S{\o}rbye, Sigrunn H and Gallego-Fern{\'a}ndez, Juan B and Zunzunegui, Mar{\'\i}a and Esquivias, M Paz and Travis, Justin MJ},
  journal={Methods in Ecology and Evolution},
  volume={4},
  number={4},
  pages={305--315},
doi = {https://doi.org/10.1111/2041-210x.12017},
url = {https://besjournals.onlinelibrary.wiley.com/doi/abs/10.1111/2041-210x.12017},
  year={2013},
  publisher={Wiley Online Library}
}

@book{krainski2018,
  title={Advanced spatial modeling with stochastic partial differential equations using R and INLA},
  author={Krainski, Elias and G{\'o}mez-Rubio, Virgilio and Bakka, Haakon and Lenzi, Amanda and Castro-Camilo, Daniela and Simpson, Daniel and Lindgren, Finn and Rue, H{\aa}vard},
  year={2018},
  publisher={Chapman and Hall/CRC}
}

@InProceedings{monteiro2022,
author="Monteiro, Andreia
and Carvalho, Maria Luc{\'i}lia
and Figueiredo, Ivone
and Sim{\~o}es, Paula
and Nat{\'a}rio, Isabel",
editor="Bispo, Regina
and Henriques-Rodrigues, L{\'i}gia
and Alpizar-Jara, Russell
and de Carvalho, Miguel",
title="Intensity-Dependent Point Processes",
booktitle="Recent Developments in Statistics and Data Science",
year="2022",
publisher="Springer International Publishing",
address="Cham",
pages="123--136",
doi= {https://doi.org/10.1007/978-3-031-12766-3_10}
}

@misc{natario2026testingpreferentialsampling,
      title={Testing Preferential Sampling}, 
      author={Isabel Natario and Andreia Monteiro},
      year={2026},
      eprint={2606.14615},
      archivePrefix={arXiv},
      primaryClass={stat.ME},
      url={https://arxiv.org/abs/2606.14615}, 
}

@article{pati2011,
  title={Bayesian geostatistical modelling with informative sampling locations},
  author={Pati, Debdeep and Reich, Brian J and Dunson, David B},
  journal={Biometrika},
  volume={98},
  number={1},
  pages={35--48},
  year={2011},
  publisher={Oxford University Press}
}

@InProceedings{paula2023,
author="Sim{\~o}es, Paula
and Carvalho, M. Luc{\'i}lia
and Figueiredo, Ivone
and Monteiro, Andreia
and Nat{\'a}rio, Isabel",
editor="Gervasi, Osvaldo
and Murgante, Beniamino
and Rocha, Ana Maria A. C.
and Garau, Chiara
and Scorza, Francesco
and Karaca, Yeliz
and Torre, Carmelo M.",
title="Black Scabbardfish Species Distribution: Geostatistical Inference Under Preferential Sampling",
booktitle="Computational Science and Its Applications -- ICCSA 2023 Workshops",
year="2023",
publisher="Springer Nature Switzerland",
address="Cham",
pages="303--314",
doi={https://doi.org/10.1007/978-3-031-37108-0_19}

}

@InProceedings{paula2022,
author="Sim{\~o}es, Paula
and Carvalho, Maria Luc{\'i}lia
and Figueiredo, Ivone
and Monteiro, Andreia
and Nat{\'a}rio, Isabel",
editor="Bispo, Regina
and Henriques-Rodrigues, L{\'i}gia
and Alpizar-Jara, Russell
and de Carvalho, Miguel",
title="Geostatistical Sampling Designs Under Preferential Sampling for Black Scabbardfish",
booktitle="Recent Developments in Statistics and Data Science",
year="2022",
publisher="Springer International Publishing",
address="Cham",
pages="137--151",
doi={https://doi.org/10.1007/978-3-031-12766-3_11}
}

@article{pennino2018,
author = {Pennino, Maria Grazia and Paradinas, Iosu and Illian, Janine B. and Muñoz, Facundo and Bellido, José María and López-Quílez, Antonio and Conesa, David},
title = {Accounting for preferential sampling in species distribution models},
journal = {Ecology and Evolution},
volume = {9},
number = {1},
pages = {653-663},
doi = {https://doi.org/10.1002/ece3.4789},
url = {https://onlinelibrary.wiley.com/doi/abs/10.1002/ece3.4789},
year = {2019}
}

@article{raeisi2020,
  title={On spatial and spatio-temporal multi-structure point process models},
  author={Raeisi, Morteza and Bonneu, Florent and Gabriel, Edith},
  journal={arXiv preprint arXiv:2003.01962},
  year={2020}
}

@article{robinsonetal2017,
AUTHOR={Robinson, Néstor M. and Nelson, Wendy A. and Costello, Mark J. and Sutherland, Judy E. and Lundquist, Carolyn J.},
TITLE={A Systematic Review of Marine-Based Species Distribution Models (SDMs) with Recommendations for Best Practice},
JOURNAL={Frontiers in Marine Science},
VOLUME={4},
YEAR={2017},
}

@article{roos2011,
  title={Sensitivity analysis in Bayesian generalized linear mixed models for binary data},
  author={Roos, Ma{\l}gorzata and Held, Leonhard},
  journal={Bayesian Analysis},
  volume={6},
  number={2},
  pages={259--278},
  year={2011},
  publisher={International Society for Bayesian Analysis}
}

@article{simpson2017,
author = {Daniel Simpson and H{\aa}vard Rue and Andrea Riebler and Thiago G. Martins and Sigrunn H. S{\o}rbye},
title = {{Penalising Model Component Complexity: A Principled, Practical Approach to Constructing Priors}},
volume = {32},
journal = {Statistical Science},
number = {1},
publisher = {Institute of Mathematical Statistics},
pages = {1 -- 28},
year = {2017},
doi = {10.1214/16-STS576},
URL = {https://doi.org/10.1214/16-STS576}
}

@article{stoyan2008,
  title={Modelling marked point patterns by intensity-marked Cox processes},
  author={Ho, Lai Ping and Stoyan, D},
  journal={Statistics \& Probability Letters},
  volume={78},
  number={10},
  pages={1194--1199},
  year={2008},
  publisher={Elsevier}
}

@article{van2020bayesian,
  title={Bayesian rank-based hypothesis testing for the rank sum test, the signed rank test, and Spearman's $\rho$},
  author={van Doorn, Johnny and Ly, Alexander and Marsman, Maarten and Wagenmakers, E-J},
  journal={Journal of Applied Statistics},
  volume={47},
  number={16},
  pages={2984--3006},
  year={2020},
  publisher={Taylor \& Francis}
}

@phdthesis{watson2020,
	title={Accounting for preferential sampling in the statistical analysis of spatio-temporal data},
	author={Watson, Joe},
	year={2020},
	school={University of British Columbia}
}

@misc{watson2020fast,
      title={A fast Monte Carlo test for preferential sampling}, 
      author={Joe Watson},
      year={2020},
      eprint={2003.01319},
      archivePrefix={arXiv},
      primaryClass={stat.ME},
      url={https://arxiv.org/abs/2003.01319}, 
}

@article{watson2021,
  title={A perceptron for detecting the preferential sampling of locations and times chosen to monitor a spatio-temporal process},
  author={Watson, Joe},
  journal={Spatial Statistics},
  volume={43},
  pages={100500},
  year={2021},
doi = {https://doi.org/10.1016/j.spasta.2021.100500},
url = {https://www.sciencedirect.com/science/article/pii/S2211675321000105},
  publisher={Elsevier}
}

@article{zidek2014,
author = {James V. Zidek and Gavin Shaddick and Carolyn G. Taylor},
title = {{Reducing estimation bias in adaptively changing monitoring networks with preferential site selection}},
volume = {8},
journal = {The Annals of Applied Statistics},
number = {3},
publisher = {Institute of Mathematical Statistics},
pages = {1640 -- 1670},
year = {2014},
doi = {10.1214/14-AOAS745},
URL = {https://doi.org/10.1214/14-AOAS745}
}

\end{document}